\documentclass[aps,nofootinbib,showpacs,preprintnumbers,twocolumn,superscriptaddress,prl]{revtex4-1}

\usepackage{amsmath,amssymb}
\usepackage{graphicx}
\usepackage{bm}
\usepackage{color}

\definecolor{michael}{rgb}{0.9,0.,0.1}

\newcommand{\E}{\ensuremath{\mathrm{e}}}
\newcommand{\I}{\ensuremath{\mathrm{i}}}

\newcommand{\ket}[1]{| #1\rangle}

\newcommand{\ev}[1]{\langle #1 \rangle}

\begin{document}

\title{Antiferromagnetism in the 
Hubbard Model on the Bernal-Stacked Honeycomb Bilayer}

\author{Thomas C. Lang}
\email{lang@physik.rwth-aachen.de}
\affiliation{Institute for Theoretical Solid State Physics, RWTH Aachen University, Aachen, Germany}
\affiliation{JARA-HPC High Performance Computing}
\affiliation{JARA-FIT Fundamentals of Future Information Technology}
\author{Zi Yang Meng}
\affiliation{Center for Computation and Technology, Louisiana State University, Baton Rouge, Louisiana 70803, USA}
\author{Michael M. Scherer}
\affiliation{Institute for Theoretical Solid State Physics, RWTH Aachen University, Aachen, Germany} 
\affiliation{JARA-FIT Fundamentals of Future Information Technology}
\author{Stefan Uebelacker}
\affiliation{Institute for Theoretical Solid State Physics, RWTH Aachen University, Aachen, Germany} 
\affiliation{JARA-FIT Fundamentals of Future Information Technology}
\author{Fakher F. Assaad}
\affiliation{Institute for Theoretical Physics and Astrophysics, University of W\"urzburg, W\"urzburg, Germany}
\author{Alejandro Muramatsu}
\affiliation{Institut f\"{u}r Theoretische Physik III, University of Stuttgart, Stuttgart, Germany}
\author{Carsten Honerkamp}
\affiliation{Institute for Theoretical Solid State Physics, RWTH Aachen University, Aachen, Germany} 
\affiliation{JARA-FIT Fundamentals of Future Information Technology}
\author{Stefan Wessel}
\affiliation{Institute for Theoretical Solid State Physics, RWTH Aachen University, Aachen, Germany}
\affiliation{JARA-HPC High Performance Computing}
\affiliation{JARA-FIT Fundamentals of Future Information Technology}

\begin{abstract}
Using a combination of quantum Monte Carlo simulations, functional renormalization group calculations and 
mean-field theory, we study the Hubbard model on the Bernal-stacked honeycomb bilayer at half-filling as a
model system for bilayer graphene. The free bands consisting of two 
Fermi points with quadratic dispersions lead to a finite density of states at the Fermi level, 
which triggers an antiferromagnetic instability that spontaneously breaks sublattice and spin 
rotational symmetry once local Coulomb repulsions are introduced. Our results reveal an inhomogeneous 
participation of the spin moments in the ordered ground state, with enhanced moments at the threefold 
coordinated sites. Furthermore, we find the antiferromagnetic ground state to be robust with respect to 
enhanced interlayer couplings and extended Coulomb interactions.
\end{abstract}

\pacs{71.27.+a,71.10.Fd,71.30.+h,73.21.Ac,75.70.Cn}

\maketitle
There is currently significant interest in understanding the electronic properties of bilayer graphene 
(BLG), in particular the ground state at the charge neutrality point. Several experimental studies 
\cite{exp01,exp02,exp03,exp04,exp05,exp06,exp07,exp08} hint to the formation of a symmetry broken state 
in BLG, but its actual nature remains ambiguous and is at the moment a highly debated topic. Symmetry 
breaking in BLG can arise due to thermal annealing-induced strain on suspended samples as well as 
external electric fields applied perpendicular to the BLG sheets. In the absence of such external 
perturbations, due to the finite density of states at the Fermi level in the free band limit, the 
electronic Coulomb interaction is expected to trigger a genuine electronic instability and drive BLG 
into a correlated ground state \cite{castroneto2009}.
Possible candidate states that have been suggested 
\cite{nilsson2006,mccann2007,hongki2008,lemonik2010,vafek2010a,nandkishore2010,zhang2010,jung2011,vafek2010b,
vafek2011,khari2011,kotov2010,scherer2012} include an (layered) antiferromagnetic (AF) state, several 
topological states such as quantum anomalous Hall, quantum spin Hall (QSH) or quantum valley Hall states, 
all of which  exhibit a finite bulk gap, as well as a gapless nematic state.  While most 
recent experiments identified a finite excitation gap of a few meV emerging in BLG at low temperatures 
\cite{exp05,exp06,exp07,exp08}, the transport data in Ref. \cite{exp04} have been interpreted towards the 
formation of a gapless, possibly nematic state. Within the currently inconclusive experimental situation, 
an AF state is considered a probable ground state \cite{scherer2012,gorbar2012,fzhang2012} among the 
(gapfull) candidates and thus worth a more detailed examination. Furthermore, the validity of approximative
approaches  need to be tested against unbiased and numerically exact results.

\begin{figure}[t!]
\centering
  \includegraphics[width=\columnwidth]{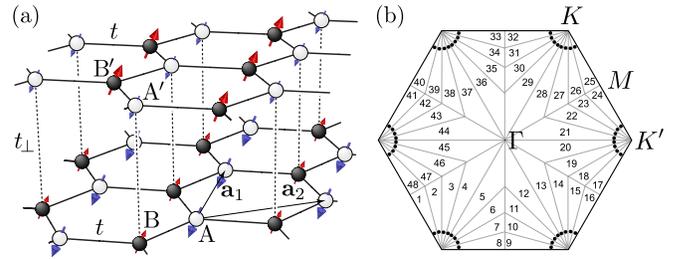}
  \caption{(a) Bernal stacking of the honeycomb bilayer with intra- (inter)layer hopping $t$ 
  ($t_{\perp}$) between the sublattices A, B and A$'$, B$'$ (A$'$, B). Within the sublattices an equal
  number of sites have a coordination number $z=3$ or 4. (b) Patching scheme of the Brillouin zone in the 
  fRG. Dots denote the momenta at which the vertex function is evaluated.
  \label{fig:lattice}}
\end{figure}

Here, we explore the nature of this possible ground state by taking screened Coulomb interactions into 
account within a tight-binding approach for BLG via a Hubbard model description of the carbon $\pi$-
electrons. In particular, since the neutrality point relates to half-filling in the Hubbard model 
description, we take the opportunity to explore possible electronic instabilities using unbiased quantum 
Monte Carlo (QMC) methods. Our
simulations are furthermore augmented by functional renormalization group (fRG) calculations~
\cite{metzner2011,scherer2012}. The fRG allows us to investigate the stability of the AF state obtained 
with QMC simulations over a broad range of the interaction strength. We find that within the AF ground
state a local spin moment's participation in the AF order anticorrelates to its lattice coordination 
number $z$, with $z=3$ or $4$ for the Bernal stacking, an effect that we show to hold over the full 
parameter range from weak to strong electronic correlations.
 
In the following we consider the Hamiltonian $H=H_{0}+H_\text{int}$, with the local 
interaction term
$H_\text{int}=U\sum_i n_{i,\uparrow}n_{i,\downarrow}$ and $n_{i,\sigma}=c^{\dagger}_{i,\sigma}
c^{\phantom{\dagger}}_{i,\sigma}$ the density operator at site $i$ for spin $\sigma$. Furthermore, 
$H_{0}$ denotes the free tight-binding model \cite{castroneto2009} containing both intralayer nearest neighbor hopping $t$ as 
well as interlayer hopping $t_\perp$, as illustrated in Fig.~1(a). For the onsite interaction $U$ and 
also a finite set of nonlocal density-density interaction parameters, {\it ab initio} calculations list 
values for graphene and graphite~\cite{wehling2011}, which have been used to explore the phase diagram of 
the honeycomb bilayer by means of the fRG approach that is also employed in the present work~
\cite{scherer2012}. It was found that 
AF order is the dominant instability for interaction parameters with a shorter range than 
those for single layer graphene. As for a Bernal-stacked bilayer, 
where screening is expected to be effective, the antiferromagnet seems 
to be a viable candidate for the ground state of BLG. In this Letter, we employed an improved fRG 
patching of the Brillouin zone into 48 sectors, cf. Fig.~1(b), in order to obtain more accurate estimates 
for the critical energy scale $\Lambda_c$ where an electronic instability emerges during the fRG flow 
while successively integrating out the high-energy modes. Details on the fRG approach have been presented 
in Ref.~\onlinecite{scherer2012}. From analyzing the structure of the resulting interaction vertex, we 
can identify the leading instability below $\Lambda_c$ for varying sets of initial coupling parameters~
\cite{metzner2011}. For a broad range of pure Hubbard interactions $U$, we 
observe a flow to strong coupling with the signature of an AF instability and an exponential dependence 
of $\Lambda_c$ on $U$ as discussed below. Up to an order of magnitude, $\Lambda_c$ can serve as estimate for 
the single particle gap $\Delta_{\text{sp}}$ in the AF state. The AF instability is robust with respect 
to variations of the band structure, in particular the interlayer coupling, which we have explicitly 
checked for $t_\perp=0.1t$ and $t_\perp=t$ (for BLG $t_\perp\approx 0.13 t$~
\cite{lmzhang2008}). We take this as further motivation for a systematic analysis of the local Hubbard 
model at half-filling. 

Our main findings result from analyzing this local Hubbard limit, where we can efficiently employ a 
projector QMC approach~\cite{AsEv08}, to perform a numerically exact evaluation of the ground state 
properties. We furthermore fix $t_{\perp}=t$; while this takes us beyond the regime of realistic 
parameters for BLG, the choice for $t_{\perp}$ allows us to reliably study electronic instabilities, due 
to the well pronounced quadratic band touching at the Fermi level~\cite{footnote1}. We performed QMC 
simulations on finite systems of linear extent $L$ (the number of sites being $N=4L^2$) for $L$ up to 
$12$ with periodic boundary conditions. Our implementation also allows for the efficient measurement of 
unequal-time correlation functions~\cite{FeAs01}. From a fit of the imaginary-time displaced Green's 
function ${G(\mathbf{q},\tau) =  \ev{\frac{1}{N}\sum_{s,\sigma} c_{\mathbf{q}s\sigma}^{\dagger}(\tau) 
c_{\mathbf{q}s\sigma}}}$ to its long-time behavior, ${\lim_{\tau\to\infty} G(\mathbf{q},\tau)\propto\E^{-
\tau\Delta_{\text{sp}(\mathbf{q})}}}$, the single-particle gap $\Delta_{\text{sp}}=\Delta_{\text{sp}}(K)$ 
can be extracted without the need of an analytical continuation. Here, $s$ labels the four orbitals per 
unit cell of the honeycomb bilayer. In order to gain information on the influence of finite size (FS) effects 
we found it useful to compare also to fRG as well as to a AF mean-field theory (MFT) decoupling of the 
interaction term $H_\text{int}$. We solved the resulting saddle-point equations self consistently on finite 
lattices and zero temperature to obtain the MFT order parameters for the sublattice magnetization and the associated
Hartree-Fock single particle gap ${\Delta_\text{sp}^\text{MFT} = \min_{\{\mathbf{k}\}}
\sqrt{\varepsilon(\mathbf{k})^2+(U m)^2}=Um}$, where ${\varepsilon(\mathbf{k})}$ the single particle 
dispersion of the free Hamiltonian $H_0$. 

\begin{figure}[tp]
\centering
   \includegraphics[width=\columnwidth]{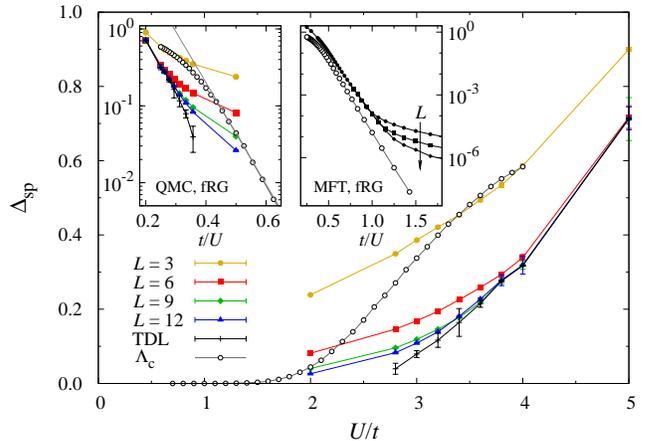}
  \caption{Single particle gap $\Delta_{\text{sp}}$ from QMC simulations for different system sizes and the finite
  size extrapolation to the TDL using a polynomial fit function, along with the fRG critical scale 
  $\Lambda_c$ as a function of the local Coulomb repulsion $U/t$. The inset on the left shows the same 
  data vs $t/U$ in a semilog scale, exhibiting an exponential onset of the gap in the large $t/U$ 
  range. The inset on the right shows the fRG data along with MFT results for system sizes $L=129$, 258, and 516.
  \label{fig:spgap}
}
\end{figure}

In Fig.~\ref{fig:spgap}, we present our results for the single particle gap $\Delta_\text{sp}$ obtained 
from QMC simulations and MFT alongside the fRG critical scale $\Lambda_c$ as a function of $U/t$. The QMC values 
exhibit a continuously increasing $\Delta_\text{sp}$ for all system sizes. The FS extrapolation 
to the thermodynamic limit (TDL) using a second order polynomial yields a continuous onset in the 
thermodynamic limit. Finite size extrapolation of the available data points deceptively suggest a finite 
critical value of $U$ for the transition from the semimetal to the Mott insulator. We will show in the 
following that this can be attributed to pronounced FS effects at low energies. The fRG critical 
scale $\Lambda_c$ (open circles) for the same parameters indeed reproduces such a continuously increasing
associated single particle gap, for all finite values of $U$. 

To identify whether the observed gap indeed is the consequence of a ${U=0^+}$ instability, we plot
$\Delta_\text{sp}$  as a function of ${t/U}$ in the insets of Fig.~\ref{fig:spgap}. The fRG data show an 
exponential opening of $\Delta_\text{sp}$ in the large ${t/U}$ regime (left inset). Accordingly, the 
FS extrapolated QMC data follow the same behavior. One can readily see that larger lattices are 
needed in order to clearly identify this exponential onset at smaller values of $U/t$. The same effect is 
observed already within the MF approach (right inset): increasingly larger system sizes allow us to identify 
the exponential opening of the single particle gap ${\Delta_\text{sp} \propto \exp(-\alpha t/U)}$. From 
our fRG data we can extract the exponent ${\alpha \approx 16}$, which is close to the Hartree-Fock value 
of ${9\pi t^2/2t_{\perp}}$ \cite{Ge97}. Deviations from this exponential behavior emerge beyond an 
intermediate coupling strength of $U/t\approx 2$ and relate to the onset of the strongly correlated 
regime. Note that within this model, the energy gap for realistic values of the Hubbard 
${U\approx 3t}$ are rather large, ${\Delta_\text{sp} \approx 0.07t \approx 200}$ meV \cite{wehling2011}.

\begin{figure}[tp]
\centering
  \includegraphics[width=\columnwidth]{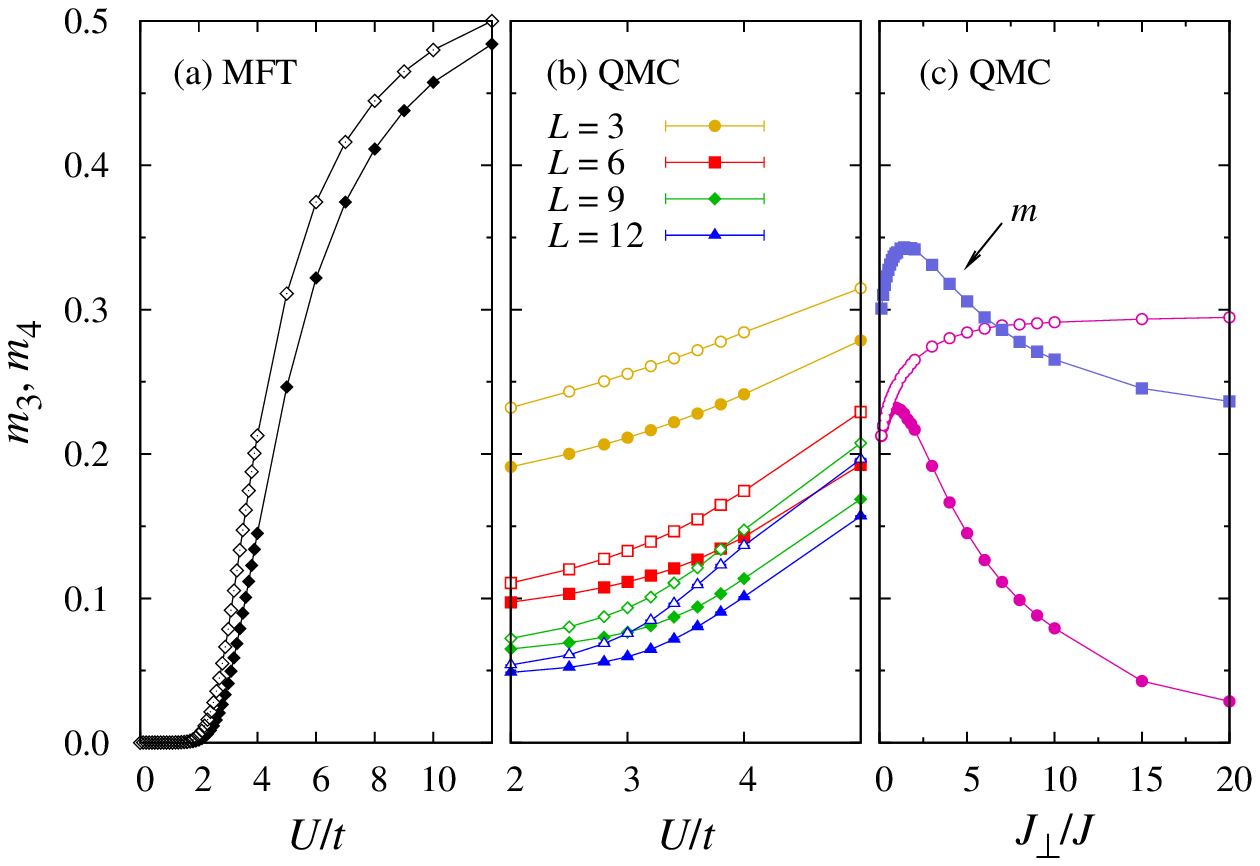}\\
  \includegraphics[width=\columnwidth]{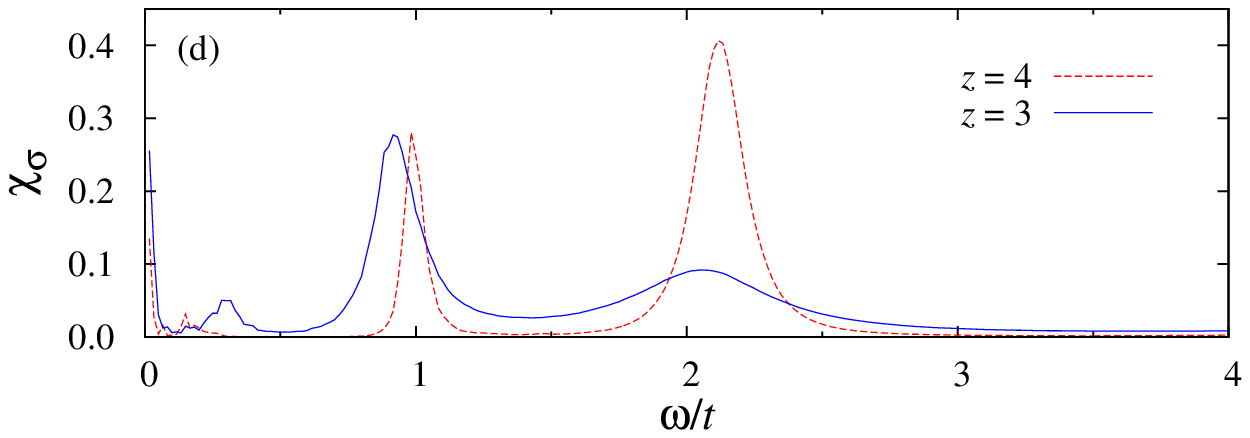}
  \caption{The sublattice magnetization vs. $U/t$ for sites with coordination number ${z=3}$ (open
  symbols) and ${z=4}$ (filled symbols) from (a) MFT in the TDL, (b) QMC simulations for different system sizes, and 
  (c) in the Heisenberg limit vs. the interlayer AF exchange coupling $J_{\perp}/J$. The data in (c) 
  result from SSE QMC simulations, after extrapolation to the TDL; also shown in (c) is the overall 
  staggered magnetization $m$. Part (d) shows the local dynamical spin susceptibility $\chi_{\sigma}(\omega)$ 
  for sites with $z=3$ and $4$ for ${L=12}$, ${U/t=4}$.
  \label{fig:mag}
}
\end{figure}

We identify the order parameter associated with the single particle gap from the low-energy vertex in the 
fRG and from corresponding correlation functions in QMC simulations, consistently, to be long-range AF order, 
correlated between both layers. To quantify this order within QMC simulations, we measure the overall staggered 
structure factor 
${
S_\text{AF}=\frac{1}{N}\sum_{i,j} \epsilon_i \epsilon_j \langle \mathbf{S}_i\cdot \mathbf{S}_j \rangle
}$
from which we obtain the mean staggered magnetization per lattice site as ${m=\sqrt{S_\text{AF}/(4L^2)}}
$. Here, ${\epsilon_i=\pm 1}$ if site $i$ belongs to the magnetic sublattice A, A$'$ (B, B$'$), as 
indicated by  the white (black) spheres in Fig.~1. In order to probe in more detail the magnetic 
correlations, we also consider the following restricted structure factors for sites with coordination 
numbers ${z=3}$ and $4$ (for a system of linear size $L$, there are ${2L^2}$ such sites each):
${
S_{\text{AF},z}=\frac{1}{2L^2}\sum_{i,j| z_i=z_j=z} \epsilon_i \epsilon_j 
\langle \mathbf{S}_i\cdot \mathbf{S}_j \rangle
}$
from which we obtain local order parameters ${m_z=\sqrt{S_{\text{AF},z}/(2L^2)}}$ for lattices sites with 
$z=3,4$. While the  overall staggered magnetization $m$ steadily increases with $U> 0$ like the single 
particle gap, we observe pronounced differences in the two sublattice magnetizations, which arise due to 
the presence of inequivalent sites in the lattice structure [cf. Fig.~\ref{fig:mag}(a)-(b)]. In 
particular, we find that sites with the higher coordination ${z=4}$ (filled symbols) exhibit a lower 
ordered moment, at odds with the usual intuition that high coordination favors more robust N\'{e}el 
order. An increase of $z$ on one sublattice will generically make the ordering more robust everywhere, 
although not uniformly so for all sublattices. The same hierarchy of magnetic moments can also be 
inferred from fRG calculations for a wider range of nonlocal interactions by comparing the relative 
strengths of the effective interactions on the different sites of a unit cell.

Similar effects of an enhanced magnetic order near low-coordinated sites have previously been observed in 
localized quantum spin models on other inhomogeneous lattice structures~\cite{JaMoWe06}. Here, the joint 
bonds on sites with coordinated number ${z=4}$ interconnect the two layers [cf. Fig.~\ref{fig:lattice}
(a)]. With increasing interaction ${U}$ and hopping $t_{\perp}$, the moments along these bonds hybridize 
between the layers and tend to form spin singlets, suppressing the participation in the long range AF 
order on these sites. In Fig.~\ref{fig:mag}(c), we consider the staggered magnetizations in the large-$U$ 
limit, wherein the model becomes a spin-only Heisenberg model, with an intralayer exchange coupling ${J=4 
t^2/U}$ and an interlayer coupling ${J_\perp=4 t_\perp^2/U}$. To study this Heisenberg limit of the 
Hubbard model, we employed the stochastic series expansion (SSE) QMC approach~\cite{sse,footnote2}, and present our 
results after an extrapolation to the TDL. We find that all three order parameters exhibit an initial 
increase upon increasing $J_{\perp}/J$. Furthermore, while $m_3$ saturates for large ${J_{\perp}/J}$, $m$ 
and $m_4$ scale to zero in the large $J_{\perp}$ limit. These results show that for all finite values of 
${J/J_{\perp}}$, the system remains antiferromagnetically ordered, but with a suppressed staggered moment 
on the ${z=4}$ sites for large ${J_{\perp}/J}$, due to the strong tendency towards forming $J_{\perp}$-
singlets along these interlayer bonds. 
If the two honeycomb lattices were stacked such that each lattice 
site would be coupled via $J_{\perp}$ to the other layer, a complete decoupling into local singlets would 
destroy the AF order beyond a finite critical value of $J_\perp/J$.

\begin{figure}[tp]
\centering
  \includegraphics[width=\columnwidth]{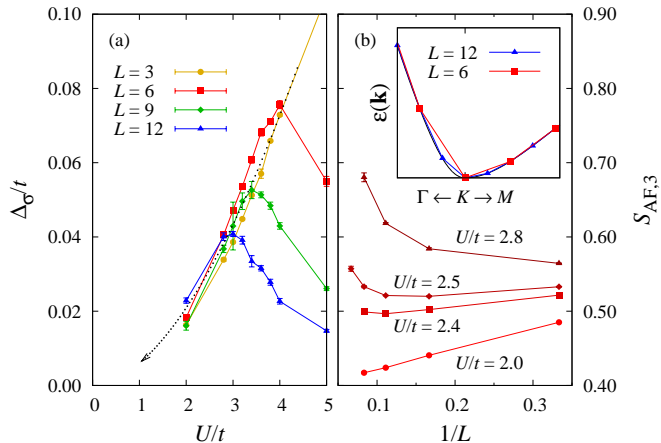}
  \caption{Finite size behavior of (a) the spin gap $\Delta_{\sigma}$ and (b) the AF structure factor
  $S_\text{AF,3}$ on lattice sites with coordination number ${z=3}$. The inset in (b) illustrates the
  discretization effects on the free  dispersion relation, a dominant source of finite size effects. 
  \label{fig:fseffects}
}
\end{figure}

Returning to the Hubbard model, the local dynamical spin 
susceptibility $\chi_{\sigma,i}(\omega) = \sum_{n} |\langle n|S^x_i|0\rangle|^2\, 
\delta(E_n-E_0-\omega)$ also exhibits defined differences depending on the local coordination, cf. the 
QMC data in Fig.~\ref{fig:mag}(d). While the peak near $\omega=0$, related to the Goldstone mode in the 
TDL, is shared by both types of sites, the $z=4$ sites exhibit a considerable shift of the residual 
spectral weight to larger energies as compared to the $z=3$ sites, in accordance with the associated 
reduced magnetic moment for $z=4$. 

To explore the global spin dynamical properties, 
Fig.~\ref{fig:fseffects}(a) shows the spin gap $\Delta_{\sigma}$ from QMC simulations
as a function of ${U/t}$ for different system sizes (we obtain ${\Delta_{\sigma}}$ from the time-displaced
spin-spin correlation function in the AF sector, ${S_{\text{AF}}(\tau) = \frac{1}{N}\sum_{i,j} 
\epsilon_i\epsilon_j \ev{\mathbf{S}_i(\tau) \cdot \mathbf{S}_j}}$). The data for finite lattices 
exhibits a pronounced peak at intermediate values of ${U/t}$; for increasingly larger lattice sizes, the 
position of this spin-gap dome shifts towards lower values of $U/t$, and its magnitude decreases, 
suggesting that in the TDL the spin gap vanishes for all values of ${U/t}$. While this is consistent with 
the emergence of Goldstone modes which originate in the spontaneous breaking of the SU(2) spin symmetry 
in the AF phase, it furthermore  illustrates the pronounced FS effects on the accessible range 
of system sizes. Similarly distinct FS corrections are also evident in the AF structure factor 
$S_{\text{AF},3}$, shown vs. $1/L$ in Fig.~\ref{fig:fseffects}(b). Consider, for example, the data for $
{U/t=2.4}$: the initial downscaling of $S_{\text{AF},3}$ on small lattice sizes would suggest a 
magnetically disordered state. However, at larger lattice sizes the scaling behavior changes, and 
eventually a strong increase of $S_{\text{AF},3}$ with system size accounts for the formation of 
long-range AF order in the TDL. This peculiar FS scaling in fact arises both in the QMC simulations and the MFT 
order parameters (not shown). The FS effects become more pronounced for smaller $t_{\perp}$. The data in 
Fig.~\ref{fig:fseffects}(b) also show that the corresponding 
crossover length scale beyond which the eventual increase of $S_{\text{AF},3}$ sets in, increases with 
decreasing values of $U/t$. Finite lattices, i.e., momentum space discretization, introduce a 
corresponding artificial gap which acts as a cutoff for correlations --- an issue which afflicts all 
finite lattice simulations. When dealing with a Fermi surface instability as in the present case, this 
generic FS effect becomes predominant on the accessible system sizes. The inset of 
Fig.~\ref{fig:fseffects}(b) illustrates this discretization of the free dispersion around the Fermi level. 
Only for sufficiently large lattices will the parabolic band be approximated to such an extent as to 
probe the TDL nonlinear low energy dispersion. 

In conclusion, we found from quantum Monte Carlo simulations, that the Hubbard model on the Bernal-stacked
honeycomb bilayer as a basic model for BLG is prone to a Fermi point instability, 
which triggers layered AF order. Characterized by their different coordination numbers, sublattice sites 
sustain different magnetic moments. This peculiar local structure of the AF state relates to its 
stability in the strong interlayer tunneling region. 
A full quenching of the magnetic moments on the $z=4$ sites emerges only in the (unrealistic) strong interlayer coupling limit
and would eventually realize the layered AF state observed within chiral two-band models for bilayer graphene. 
Functional renormalization group calculations support 
the stability of the AF state and its moment distribution over a wide range of coupling parameters.
In case the experimental support for a gapped state in BLG 
will be substantiated, it might be interesting to search for such an inhomogeneous AF state by local 
moment-sensitive probes such as magnetic scanning tunneling microscopy. 

\begin{acknowledgments}
We thank S.~Bl\"ugel, M.~J.~Schmidt, O.~Vafek, T.~Wehling, and F.~Zhang for valuable discussions. 
This research was supported in part by the DFG research units FOR~723, 912, and 1162 and the NSF 
EPSCoR Cooperative Agreement No. EPS-1003897 with additional support from the Louisiana Board of
Regents. Furthermore, we acknowledge the JSC J\"ulich and the HLRS Stuttgart for the allocation of CPU time.
\end{acknowledgments}


\begin{thebibliography}{99}
\bibitem{footnote1}
We use a trial wave function ${\ket{\Psi_\text{T}} = \prod_{\sigma} \ket{\Psi_\text{T}}_{\sigma}}$, 
where $\ket{\Psi_\text{T}}_{\sigma}$ is the ground state of the single particle Hamiltonian ${H_{\sigma}
^0 = -t \sum_{\langle i,j\rangle} c^{\dagger}_{i\sigma} c_{j\sigma} \exp\big(\frac{2\pi\I}{\Phi_{0}}
\int_{\mathbf{r}_i}^{\mathbf{r}_j}\text{d}{\boldsymbol\ell}\cdot\mathbf{A}\big) + \mbox{H.c.} }$ in the spin $
\sigma$ Hilbert subspace, where ${\Phi_{0} = h e/c}$ denotes the flux quantum, and $\mathbf{r}_i$ the 
position of lattice site $i$. The flux ${\Phi/\Phi_0}=10^{-5} $ is chosen sufficiently small such that it
is negligible within the achieved numerical accuracy to lift the ground state degeneracy in 
${|\Psi_\text{T}\rangle}$. We employed a projection parameter ${\Theta t = 
50}$ and an imaginary time discretization of ${\Delta\tau t = 0.025}$.
\bibitem{footnote2}
The SSE results have been obtained from finite size simulations with $L=6$, 12, 18, 24 at temperatures (couplings) $\beta/J=6 L$ representative for the ground state and their consecutive linear extrapolation to the TDL.
\end{thebibliography}
\end{document}